\documentclass[12pt,preprint]{aastex}
\usepackage[]{natbib}

\slugcomment{Not to appear in Nonlearned J., 45.}

\shorttitle{Total linear polarization in the OH maser W75N}
\shortauthors{Slysh et al.}

\begin{document}


\title{Total linear polarization in the OH maser W75N:\\
VLBA polarization structure.}


\author{V. I. Slysh}
\affil{Astro Space Center, Lebedev Physical Institute, Profsoyuznaya
  84/32, 117810 Moscow, Russia}
\email{vslysh@asc.rssi.ru}  

\author{V. Migenes}
\affil{University of Guanajuato, Department of Astronomy, Apdo Postal 144, Guanojuato,
  CP 36000, GTO Mexico}
%\email{vmigenes@astro.ugto.mx}

\author{I. E. Val'tts}
\affil{Astro Space Center, Lebedev Physical Institute, Profsoyuznaya
  84/32, 117810 Moscow, Russia}

\author{S. Yu. Lyubchenko}
\affil{Astro Space Center, Lebedev Physical Institute, Profsoyuznaya
  84/32, 117810 Moscow, Russia}
  
\author{S. Horiuchi}
\affil{Astro Space Center, Lebedev Physical Institute, Profsoyuznaya
  84/32, 117810 Moscow, Russia}
  
\author{V. I. Altunin}
\affil{Jet Propulsion Laboratory, 4800 Oak Grove Dr.,
	Pasadena, CA 91109, USA}
  
\author{E. B. Fomalont}
\affil{National Radio Astronomy Observatory, 520 Edgemont Rd., Charlottesville,      
	VA22903, USA}

\and

\author{M. Inoue}
\affil{Nobeyama Radio Observatory, Nobeyama, Minamisaku,
	Nagano 384-13, Japan} 


\begin{abstract}
W75N is a star-forming region containing various ultracompact HII regions 
and OH, H$_2$O, and CH$_3$OH maser emission. Our VLBA map shows that the OH 
masers are located in a thin disk rotating around
an O-star which is the exciting star of the ultracompact HII region VLA1.
A separate set of maser spots is connected with 
the ultracompact HII region VLA2. The radial velocity of OH maser spots
varies across the disk from 3.7~km~s$^{-1}$ to 10.9~km~s$^{-1}$. 
The diameter of the
disk is 4000~A.U. All maser spots are strongly polarized. This are the first
OH masers showing nearly 100 per cent linear polarization in several spots. 
Two maser spots seem to be Zeeman pairs corresponding to a magnetic field of 
5.2~mgauss and 7.7~mgauss, and in one case we tentatively found a Zeeman pair consisting
of two linearly polarized components. 
The linearly polarized maser spots are shown to be 
$\sigma$-components which is 
the case of the magnetic field being perpendicular to the line of sight. 
The direction of the magnetic field as determined from  
linearly polarized spots is perpendicular to the plane of the disk, although
the galactic Faraday rotation may significantly affect this conclusion.
\end{abstract}


\section{Introduction}

OH maser emission is strongly polarized, usually in a circular mode,
although an admixture of linear polarization was found in some masers,
which resulted in the elliptical polarization mode. The importance of 
the polarization properties of OH masers lies in the physics of the 
maser excitation and saturation properties. Current theories of OH 
maser emission polarization attribute it to the magnetic field in the 
emission region, of the order of several milligauss. This is three 
orders of magnitude larger than the general galactic magnetic field, 
but is consistent with estimates of the maser emission region density 
of $10^6-10^7$ ~cm$^{-3}$ as originating from the compression of the 
interstellar matter and magnetic field. Determination of the magnitude
and direction of the magnetic field will help to better understand the role
of the magnetic field in the formation of circumstellar disks or shocks
at the interface between compact HII regions and the ambient molecular cloud.

The polarization properties of OH maser emission were predicted by 
theoretical models on the basis of Zeeman splitting \citep*{goldreich73}.
If the Zeeman splitting exceeds the line width
(B$>$0.5 milligauss), three 100 per cent polarized components are predicted:
unshifted $\pi$--component linearly polarized along the projected 
direction of the magnetic field, and two 100 per cent elliptically polarized 
$\sigma$--components shifted in frequency to both sides of the $\pi$--component.
The relative intensity of the $\pi$--component and the degree of 
ellipticity depend on the angle $\theta$ between the line of sight and
the magnetic field direction. If $\theta=\pi/2$, the $\pi$--component 
is at maximum, and $\sigma$--components are linearly polarized 
perpendicular to the magnetic field direction. When $\theta=0$, the 
intensity of the $\pi$--component is  zero, and $\sigma$--components 
are 100 per cent circularly polarized in opposite directions. When 
$0<\theta<\pi/2$, the $\pi$--component is linearly polarized along the
magnetic field direction and is of intermediate intensity, and 
$\sigma$--components are elliptically polarized, with the 
major axis of the ellipsis perpendicular to the magnetic field. Such 
a picture has never been observed in OH masers in full details.
Typically, individual maser spots were found to be highly circularly 
polarized only in one sense. In some cases opposite circular polarization
was found to come from the same maser spot and was interpreted as a 
Zeeman pair. Linear polarization was rarely observed, and usually was
found as part of the elliptically polarized emission. The best-studied
region of OH maser polarized emission is W3(OH). \citet{garcia88}
found several dozens of circularly polarized components, with 16 being
elliptically polarized. Only three components had linear polarization
greater than circular polarization, with a maximum fractional linear
polarization of 46 per cent. Five Zeeman pairs were identified. 
\citet{garcia88} came to the conclusion that no features were detected that might
have been identified as $\pi$--components. Most of the details were 
single, circularly polarized features. In a similar polarization study
of G35.02$-$0.74N with a lower angular resolution \citet{hutawarakorn99}
found 25 circularly polarized spectral features, with four
Zeeman pairs and five features in which linear polarization was also 
present, with a maximum fractional polarization of 63 per cent. They were 
considered as $\sigma$--components. 

W75N is another well-studied OH maser source. It is associated with a 
star--forming region and an ultracompact HII-region, at a distance 2~kpc 
\citep{habing74}.
VLBI maps of the OH main line, 1665~MHz, show nine components
in the elongated region 1.5$''$ in extent \citep{haschick81}. From 
a single antenna polarization study \citet{haschick81} were able to 
determine polarization of some maser spots in their VLBI map, and 
suggested two Zeeman pairs with opposite circular polarization. A true
high angular resolution polarization study of W75N was conducted with
MERLIN by \citet{baart86}. They also found several Zeeman pairs 
of oppositely circularly polarized spectral features coming from the 
same position within measurement errors. In addition they suggested that 
there may be seven linearly polarized components present. This suggestion
was based on the presence of opposite circularly polarized pairs of 
spectral features coming from the same position and having the same 
radial velocity, in contrast with Zeeman pairs having different radial
velocities. No cross-correlation measurements were done in this study,
which are needed to measure linear polarization.

The prevailance of circular polarization and a scarcity of linear polarization
is a long standing problem in maser polarization theory.
\citet*{goldreich73} have proposed that the Faraday rotation in the emission 
region might destroy linear polarization, and only circular polarization
remains. If the linear polarization suggested by \citet{baart86} is 
real, it might be used for testing polarization models. The polarization 
structure of OH masers may be smeared by the lack of spectral and 
angular resolution when several adjacent features are mixed. \citet{baart86}
made their study with the angular resolution of 280~mas and 
spectral resolution of 0.3~km\,s$^{-1}$, which is insufficient as will be 
seen from our results.

In this paper we present a new polarization study of W75N with much
higher angular and spectral resolution, and  with a full polarization 
analysis making use of all Stokes parameters. This made it possible 
to exclude any contamination of spectral features by a contribution 
from other spectral features. In addition, for the first time, polarization
mapping of W75N in another main line OH transition, 1667~MHz, has been 
performed. 
          
\section{Observations and data reduction}

The observations were made on 1998 July 1 as a part of the survey of compact
OH masers suitable for further observations with the space ground 
interferometer \citep{migenes01}. We used the VLBA in a snap-shot mode, 
observing every source with a 
5 minute scan. The synthesized beam of the array at the OH 
frequency, 1665/1667 MHz, was elliptical, 12$\times$4~mas 
with the position angle of the major axis $-$5$^\circ$,
which corresponds to the size and orientation of the VLBA at the
moment of the snap-shot observation.
The observations were 
carried out at all four OH line frequencies, in two circular 
polarizations, with the bandwidth of 125 kHz (25~km\,s$^{-1}$) divided in 
128 spectral channels at each OH frequency which provided a spectral resolution
of 0.176~km\,s$^{-1}$. The correlation was done with the NRAO VLBA
correlator in Socorro in the full cross-correlation mode providing 
four correlated complex spectra RR, LL, RL and LR for each pair of 
antennas. These outputs can be used for determination of Stokes 
parameters in a standard way:

\[
I=\frac{1}{2}(RR+LL)\hspace{15mm}
V=\frac{1}{2}(RR-LL)
\]
\begin{equation}
Q=\frac{1}{2}(RL+LR)\hspace{15mm}  
U=\frac{i}{2}(LR-RL)
\end{equation}

The amplitude calibration was performed using the system noise temperature 
and gain curves provided by VLBA Operations. Continuum sources 
3C273 and OQ208 were used for the band pass and polarization 
calibration. Fig.~1a,b show images of 3C273 and OQ208 in I (contours) 
and linearly polarized intensity (vectors). The vectors
shown in Fig.~1a correspond to a fractional linear polarization
$m_L$=1.4 per cent;
the circular polarization measured on 3C273 is $m_C$=1.8 per cent.
No calibration of the position angle of the linear polarization
has been done. In principle, one could calibrate the position angle by 
rotating the measured linear polarization vector of 3C273 to the 
published value (e.g. \citep{conway93}) and applying this correction
to W75N OH maser data. However, we considered this to be of little
usefullness because observations of W75N have been performed 7.5 hours
after observations of 3C273, and during that time interval ionospheric
Faraday rotation could change significantly direction of the polarization
vector. Moreover, in view of the uncertain correction for the Galactic
Faraday rotation (Section 4.2) the absolute position angle of the linear
polarization remains unknown, and in what follows we give only relative
position angles which are rotated by an unknown amount. 
For OQ~208 linear polarization vectors correspond to $m_L\leq$0.1 
per cent they are distributed over the entire  field and  
are caused by noise; the circular polarization is also $m_C$=1.8 per cent.
OQ208 is known to be unpolarized \citep{stanghelini98}, while
the 18~cm linear polarization of the core of 3C273
shown on Fig~1a is $m_L$=1.8 -- 2.15 per cent \citep{conway93},
which is close to our result on Fig.~1a taking into account the crudeness
of the snap-shot measurements. In any case it would be safe to disregard
any linear polarization below 1 per cent. As for the circular polarization 
which 
is known to be absent in these two sources, the consistency of the two 
measurements means that there is unaccounted instrumental polarization 
in a sense that the intensity of the right circular polarization is 
systematically larger than the intensity of the left circular polarization.
In presenting the OH maser data we made a correction for the instrumental 
polarization by subtracting 1.8 per cent from all fractional polarization 
values of $m_C$. These corrections are small compared to the fractional 
polarization measured for OH spectral features.

The calibration was performed by the self-calibration of the reference
spectral feature using AIPS task "FRING". The reference feature was chosen 
such as to allow calibration of both RCP and LCP channels of the receiver.
This is possible if a spectral feature emits strongly enough in both modes
of polarization from the same position. To satisfy these requirements 
the feature must be either unpolarized, or linearly polarized. Since for OH 
masers
unpolarized features are rare, we looked for linearly polarized
features. Fig.~2 (upper) shows  the cross-polarized spectrum RL on the 
baseline 
Kitt Peak -- Pie Town. The non-zero features in this cross-polarization 
spectrum (i.e. a correlated spectrum from a pair of antennas with opposite
circular polarizations) are due to the linear polarization. There are several
linearly polarized features in the spectrum which can be taken as a reference.
We took the feature at 5.65~km\,s$^{-1}$ which showed the largest correlated 
flux on the longest baseline, Mauna Kea -- Saint Croix, and could be used for 
calibrating all the antennas of the array.

The calibration of the array was performed by means of fringe fitting
to the reference feature. A correction to the position of the phase center
adopted at the correlator was made using the measured absolute position of 
the reference feature. The absolute position was measured by the 
fringe rate method. Since the position offset of different spectral features
in W75N is known to be quite large compared to the size of the map, the 
mapping was carried out in two steps. First, an approximate position of all 
spectral features relative to the reference feature was determined with the 
fringe rate method. Then maps of spectral features were constructed centered
on these approximate positions. From the maps relative positions and angular
dimensions were determined by fitting two-dimensional Gaussians and 
deconvolving them with the beam. Separate maps of spectral features were 
obtained in all Stokes parameters: I, Q, U and V. From the first three
Stokes parameters polarization maps were constructed, which show total 
intensity I contours and linear polarization vectors. 

The 1667~MHz data were considered as independent from the 1665~MHz data and 
were calibrated separately, using the feature at the radial velocity of 
9.8~km\,s$^{-1}$ (Fig.2 lower) as a reference. The absolute position of the 
1667~MHz reference feature was
determined by the fringe rate method independently from the 1665~MHz 
measurements, with a lower accuracy than the relative position measurement 
accuracy.

\section{Results}

\subsection{Spectrum}

The cross-correlated circularly polarized spectra of the 1665 MHz line are 
shown in Fig.~3.  If compared with spectra taken in 1975 by \citet{davies77}
one can see many changes. The strongest component is the 
RCP feature near 12~km\,s$^{-1}$ with the same flux density, but with the 
radial velocity shifted to 12.45~km\,s$^{-1}$. A nearby weaker component at 
13~km\,s$^{-1}$  has disappeared since 1975. Two new RCP features have 
appeared, at 0.65~km\,s$^{-1}$ and 3.0~km\,s$^{-1}$. In the LCP spectrum a new 
component at 9.4~km\,s$^{-1}$ has appeared, as well as a component at 
0.65~km\,s$^{-1}$. The 1667~MHz spectrum shows
less spectral features, and they occupy a smaller velocity range. The total 
intensity cross-correlated spectra at 1665~MHz and 1667~MHz are shown in 
Fig. 4a,b. There seems to be no one-to-one correspondence between the 
spectral features in the two main OH lines. 
No emission features were found in 
the OH satellite lines at 1612~MHz and 1720~MHz.

\subsection{Absolute position}

The absolute position of the 1665~MHz reference feature is given in the 
footnote to Table~1. It is a factor of 3 more accurate than the previously 
measured absolute position \citep{haschick81}, but both measurements give 
coinciding positions within the combined error box. The absolute position of 
the strongest 1667~MHz feature at 9.8~km\,s$^{-1}$ was measured independently, 
 and was found to be shifted in the sky from the 1665~MHz reference feature 
by -- 175$\pm$69~mas in right acsension (offsets in right ascension mean 
R.A.$\times$cos(Dec)) and by -- 661$\pm$42~mas in declination.

\subsection{Distribution of maser spots}

Eleven maser spots were identified at the 1665~MHz transition and four at 
the 1667~MHz. In Table~1 we present the relevant parameters for the maser 
spots determined from their spectra and from polarization maps: LSR radial 
velocity, Stokes parameters
(I, V, Q, U), percentage of circular ($m_C$) and linear ($m_L$)
polarization, linear polarization position angle $\chi$, offsets in R.A. 
and Dec. from the reference position, deconvolved angular dimensions of
the fitted elliptical Gaussians and their position angle $\Theta$. The error in the 
relative positions is less than 1~mas. The position error of the 
1667~MHz spot L is much larger since it was determined by fringe rate method,
but the position errors of the rest of the 1667~MHz spots relative to spot L
is less than 1~mas. 
The position of the maser spots is shown on Fig.~5. Also shown are the 
ultracompact HII~regions VLA1 and VLA2 \citep{torrelles97}. Most of the 
spots lie on an almost straight line of 1.6~$''$ (3200 A.U.) in extent, 
slightly 
displaced from VLA1. The 1667~MHz spots do not coincide with any of 1665~MHz
spots, however they are located in the same general area. The radial velocity 
varies along the line  from 3.8~km\,s$^{-1}$ to 10.9~km\,s$^{-1}$ 
(this is the mean radial velocity of features A and B, which come from the
same position as a Zeeman pair), moving from North to South. Two 
separate spots, J and K are located near the second ultracompact HII~region 
VLA2.  The map closely resembles the MERLIN map by \citet{baart86}. All 
our 1665~MHz spots can be found on the \citet{baart86} map, except for 
spot K at 0.65~km\,s$^{-1}$. There are several weak spots on the MERLIN map 
which are not present on our map but this can be due to our poorer 
sensitivity.  The 1667~MHz map has been built for the first time, and no 
comparison with earlier maps is possible. 
Also, a good correspondence is found with the VLBI map by \citet{haschick81}.
The only significant difference between our map and those of  
\citet{baart86} and \citet{haschick81} is the shifted position of spot 
A (12.45~km\,s$^{-1}$) by 140~mas; note, that the radial velocity of 
this feature has also changed by 0.45~km\,s$^{-1}$. One can not be sure that 
feature A is identical to the 12~km\,s$^{-1}$ feature in earlier maps, it is 
possible that the 12~km\,s$^{-1}$ feature has disappeared, and that feature A 
is a new feature which has flared up not far from the 12~km\,s$^{-1}$ 
position.  On the other hand the identification of feature A with the 
12~km\,s$^{-1}$ feature is supported 
by its dominance in both spectra and by the equality of flux densities. 
            
\subsection{Spot size and brightness temperature}

All maser spots have been partially resolved. From Table~1 one can see that 
the largest spots are spot K (20~mas$\times$ 12~mas) and spot 
N (14~mas$\times$ 3~mas), but they may consist of several 
smaller spots which we can not resolve. 
The rest of the spots can be fitted with Gaussians with the major axis from
4.4 to 10.7~mas, and the minor axis from 0.7  to 4.6~mas. 
Considering that the beam size is 12 $\times$ 4~mas, 0.7~mas 
must be regarded only as a formal fitting solution. The real extent of the 
minor axis can be much less than the indicated, so it should be considered 
as an upper limit. The brightness temperature of one of the brightest 
1665~MHz spots, F (reference feature), with the integrated flux density 
of 18.1~Jy is $1.6\times$10$^{12}$K. The brightest 1667~MHz spot is L, with 
the integrated flux density 31.9~Jy, and it has a brightness temperature of 
1.9$\times$10$^{12}$~K. These must be regarded as lower limits since
most of the spots are probably not resolved along the minor axis. 
All the mapped spots are elongated, with the major to minor axis ratio 
3 or higher. 
If the size of the maser spots is not
intrinsic and is caused by the scattering in the interstellar medium, then
the elongation of the spots may be a result of the anisotropic scattering.

\subsection{Polarization of maser spots}

All maser spots given in Table~1 were imaged in all Stokes parameters.
This made it possible to determine full polarization properties of the maser 
spots. The percentage of circular polarization given in Table~1 was 
calculated as 

\begin{equation}   
\qquad m_C=100\frac{V}{I}
\end{equation}   
Positive $V$ corresponds to the right circular polarization. Percentage 
of the linear polarization was calculated as:

\begin{equation}   
\qquad m_L=100\frac{\sqrt{Q^2+U^2}}{I}                  
\end{equation}   
and the position angle of the electric vector of linear polarization was
calculated as 

\begin{equation}   
\qquad \chi=\frac{1}{2}\arctan\frac{U}{Q}
\end{equation}   
with positive direction East of North.
$I$, $Q$, $U$, $V$ are Stokes parameters at the maximum of the corresponding 
spot maps (Jy/beam). In some cases the percentage of the polarization exceeds
100. This may be due to a difference in the position of maxima in different 
Stokes parameters, as well as to errors of intensity measurements. 

From Table~1 it is evident that the maser spots are highly polarized,
typically 100 per cent polarized. There are 2~types of the maser spots. The 
first type is strongly linearly polarized spots with $m_L>$70 per cent
(E,F,G,H,J,K). Polarization of the spots in this group is not purely 
linear, it has an admixture of small circular polarization, making the 
polarization elliptical. The percentage of circular polarization varies 
from less than 0.3 per cent to 80 per cent. \citet{baart86} suggested 
several candidates as linearly polarized components, based on the small 
difference
in position and equal velocities of several oppositely circularly 
polarized pairs. Now, with the full polarization analysis and a much 
higher position accuracy we can confirm or reject  \citet{baart86}
candidates. We confirm five of seven MERLIN candidates: J=A,A; G=F,C;
F=G,D; E=H,E; A,B=N,K. The former is the spot designation in Table~1,
and the latter are designations by Baart et al. (1986). Two other 
candidates, D,B and M,J were not present in our data. The spot A,B=N,K is
predominantly circularly polarized, with linear polarization of only 9.3 
per cent.  There is a pair of oppositely polarized spectral features at the
radial velocity 9.35~km\,s$^{-1}$, which looks like a linearly polarized 
feature suitable for the calibration; but it is absent in the cross-polarized 
spectrum in Fig.2a which means that the left-hand and right-hand polarized 
emission come from different positions, which has been confirmed by the 
subsequent mapping (features B and C in table 1). This is a chance
coincidence of radial velocities of two independent maser spots with opposite
circular polarization which could be mixed with linear polarization in the 
absence of the full polarization analysis and high angular resolution 
mapping. In the 1667~MHz transition there is only one linearly polarized 
spot, with a percentage of linear polarization of 41.6 per cent, and with 
a circular polarization fraction of 23 per cent. 

The position angle of the linear polarization vector is in the range of 
$-23^\circ$ to $23^\circ$ for the spots in this group. The mean 
position angle of the linear polarization vector is $3.9^\circ\pm17^\circ$ 
(Fig.~6). Fig.~7 shows polarized and total intensity images
of several linearly polarized spots. 

The other type of maser spots is strongly circularly polarized, with an
admixture of linear polarization (spots A--D,I,M,N,O). The percentage of 
the circular polarization in this group is close to 100 per cent, and the 
fractional linear polarization is of less than 5 per cent to 17.9 per cent, 
and only spot L has moderate linear and circular polarization.

\subsection{Identification of Zeeman pairs}

The 1665~MHZ emission from spots A and B, and from spots M and O at 1667~MHz 
come from the same positions, with the position difference 0.8~mas 
and 1.1~mas, respectively, which is much less than the size of the 
beam and the size of the spots. 
The polarization is oppositely circular, and the velocity difference
is 3.07~km\,s$^{-1}$ and 2.7~km\,s$^{-1}$, respectively.  Based on these 
properties one can identify these two pairs as $\sigma{\pm}$-components of 
the Zeeman pattern, with the magnetic field 5.2~mgauss and 7.7~mgauss, 
respectively (positive sign indicates field direction pointing away from the 
Earth). Both pairs are elliptically polarized, with nearly the same position
angles of the polarization vector, exactly as required for 
$\sigma$--components. Nevertheless there are distinctions from the theoretical
Zeeman pattern. The first is the inequality of the intensities of right-hand 
and left-hand polarized emission. The intensity ratio of circularly polarized 
components is 5 for the A,B pair, and 1.8 for the M,O pair. Linearly 
polarized intensities are also different, the ratio being 2.2 for the A,B 
pair and 1.2 for the M,O pair.
Another important deviation from the theoretical Zeeman pattern is the  
complete absence of $\pi$--components. In a Zeeman pair the $\pi$--component
must be located at the mean radial velocity of the pair. We conducted an 
intense search for $\pi$--components for the two Zeeman pairs: A,B at the 
radial velocity of 10.9~km\,s$^{-1}$ and for M,O at the radial velocity of 
7.3~km\,s$^{-1}$, and did not find any emission above the noise level. Also, 
a $\pi$--component must be only linearly polarized, and we found that every 
maser spot is elliptically polarized. The only spot for which the circular 
polarization was not detected is spot E. Therefore we conclude that 
$\pi$--components are not present in the Zeeman spectrum of OH masers in W75N. 
\citet{baart86} suggested several Zeeman pairs, none of which are  
confirmed here. For the pair with radial velocities of 3.8~km\,s$^{-1}$ (LCP) 
and 6.5~km\,s$^{-1}$ (RCP) suggested by  \citet{baart86} we confirm only 
the LCP-component, with LCP/RCP ratio above 28; the LCP/RCP ratio in  \citet{baart86}
is 5. In addition, for the other Zeeman pair suggested by  \citet{baart86}
with radial velocity of 4.8~km\,s$^{-1}$ (LCP) and 
9.6~km\,s$^{-1}$ (RCP) we could not find the LCP counterpart, with the 
RCP/LCP ratio of more than 350 while  \citet{baart86} data suggest 
RCP/LCP=12. 

\citet{haschick81} suggested two possible Zeeman pairs which are not
present in our data. The discrepancy between our data and earlier results 
can be attributed to the the variability of the Zeeman component intensity.
We never see equal intensity of $\sigma$--components, and the intensity
ratio can vary in a very large range. If the $\sigma$--component intensity 
ratio is very large, the weaker component becomes unobservable. This may  
explain the presence of single circularly polarized spectral features 
in OH masers.        

\subsection{Polarization status}

On Fig.~8 we show the percentage of circular polarization plotted against
the percentage of linear polarization for all spots. The spots are labelled
by letters as in Table~1. The dashed line shows where completely polarized
features could appear on the diagram. The majority of the spots are located
close to the 100 per cent polarization line. A notable exception is spot L 
(1667~MHz transition) which is 47 per cent polarized. There is evidence for  
a correlation in that predominantly circularly polarized spots are either 
located in the upper and lower left corners of the diagram, and predominantly 
linearly polarized spots are located in the middle right of the diagram. 
The numbers on the dashed line indicate an angle $\theta$ between the line 
of sight and the magnetic field direction, according to equations

\begin{equation}
m_L=100\frac{\sin^2\theta}{1+\cos^2\theta}\hspace{15mm}
m_C=\pm100\frac{\cos\theta}{1+\cos^2\theta}  
\end{equation}

These equations describe polarization of the $\sigma$--components in the 
theory by  \citet*{goldreich73} for the case when the Zeeman
splitting is larger than the linewidth  \citep[see also][]{elitzur96}. Most 
of the maser spots occupy two regions on the diagram: 
$\theta <30^\circ$ and $80^\circ<\theta<90^\circ$. If all the spots are 
$\sigma$--components it means that the line of sight is either parallel or 
perpendicular to the magnetic field direction, and no intermediate cases exist.
The predominantly circularly polarized spots are almost certainly $\sigma$--components,
although frequently without the second member of the pair.
The predominantly linearly polarized spots are also $\sigma$--components 
with $\theta\approx$90$^\circ$, or $\pi$--components. A correct attribution is 
important for the determination of the direction of the magnetic field:
it is parallel to E-vector for $\pi$--components, and perpendicular to 
E-vector for $\sigma$--components. As mentioned in Section 3.6 the presence 
of the small circular polarization makes these components elliptically 
polarized, and therefore they must be $\sigma$--components, since 
$\pi$--components can not have a circular polarization contribution. 

The only linearly polarized spot, L at 1667~MHz, is definitely not 100 per 
cent polarized and is an exception among all maser spots in W75N. In the model
of the weak magnetic field the percentage of linear polarization for 
$\sin^2\theta>1/3$ is 

\begin{equation}
m_L=100\frac{3\sin^2\theta-2}{3\sin^2\theta} 
\end{equation}

\citep*{goldreich73}. For this spot $m_L= 41.6$ per cent, and 
corresponds to $\theta=43.4^\circ$. The percentage of circular polarization 
varies across the line profile (Fig.~9) changing the sign as predicted by the
theory for the case of small Zeeman splitting \citep{elitzur96}. However the curve
is shifted from the line centre and is strongly distorted compared to the 
theoretical one, probably because of the non-linear mode competition in 
the maser. The linear polarization increases to the red-side of the profile,
while the circular polarization is larger at the blue side.

\section{Discussion}

\subsection{Linear polarization}

The most important result of this study is the discovery of several linearly
polarized maser spots with almost 100 per cent polarization. The mere 
existence 
of linear polarization in these spots and in the circularly polarized
spots means that the Faraday rotation proposed by \citet*{goldreich73}
as a mechanism for the elimination of $\pi$--components is 
not important in W75N OH maser. The linearly polarized components are
$\sigma$--components of the Zeeman pattern.
This conclusion can be tested if the second $\sigma$--component of the Zeeman pair of linearly 
polarized emission
will be found, shifted in radial velocity by an appropriate amount. We tentatively found
the Zeeman counterpart F1 to the linearly polarized feature F. 
It is shifted in the radial velocity by
6.3~km\,s$^{-1}$, corresponding to the magnetic field strength 10.7 milligauss, and
is almost fully linearly polarized, with the position angle close to the
position angle of spot F. This is consistent with the $\sigma$--component interpretation of spots
F and F1. However, this conclusion has to be confirmed with a more sensitive measurements, 
since
the detection of the component F1 is tentative. Its intensity is only about 0.003 of intensity of
component F, and it was detected on the edge of the spectral band of the receiver, in fact, in
the last 128th channel. This is not surprising since the Zeeman pairs were found in W75N, with
unequal intensity of the components. The linearly polarized Zeeman counterpairs
might be too weak for a detection. The difference in intensity of Zeeman $\sigma$--components
seems to be a widespread property of OH masers, and is probably due to a gradient of the radial
velocity and magnetic field \citep{cook66}. The absence of $\pi$--component is more difficult to 
understand.
As mentioned above internal Faraday rotation can not be responsible for this.
Maser model calculation by \citet*{gray95} produce both $\pi$  and $\sigma$--components, 
with
$\sigma$--components dominating over $\pi$--components in the range of angles
$\theta=0-55^\circ$, where $\sigma$--components are circularly or elliptically polarized. For
$\theta=55-90^\circ$, $\pi$--components which are linearly polarized dominate.
Saturation effects of competitive gain strongly reduce a weaker component, that is
$\pi$--component for $\theta=0-55^\circ$, and $\sigma$--component 
for $\theta=55-90^\circ$.
\citet*{gray95} note that the suppression of $\sigma$--component is much less
efficient, than the suppression of $\pi$ by $\sigma$--components.
The model of \citet*{gray95} is in conflict with our conclusion that
linearly polarized spots in W75N are emitting $\sigma$--components with $\theta$
close to $90^\circ$, while the model predicts dominance of the linearly polarized
$\pi$--components for $\theta$ near $90^\circ$. Our identification of the linearly
polarized spots as $\sigma$--components was based on the presence of some circularly
polarized emission in all spots except spot E. This seems to be a very firm
argument in favor of $\sigma$--components unless an explanation for the 
circular polarization contribution to the $\pi$--components is found.

There is an alternative interpretation of the observed linear polarization as a result
of the saturated maser amplification with a weak magnetic field, when
the Zeeman splitting is less than the bandwidth. For $\sin^2\theta\le1/3$,
or $\theta\le35.5^\circ$ there will be a single 100 per cent linearly polarized emission line
with E-vector parallel to the magnetic field direction (Case 2a in \citet*{goldreich73}).
For $\theta>35.5^\circ$ the linear polarization will be lower, and the 
circular polarization appears 
in the line wings (see Section 3.7). This model could explain observed linearly polarized emission in 
W75N
but the observed admixture of circular polarization remains unaccounted, the same as for
the $\pi$--components
in the Zeeman splitting. Also this model limits the magnetic field to about 0.5 milligauss which is 
much less
than was deduced for spots A, B, F, F1 and M, O in Section 3.6. More likely this explanation could 
be applied
to the observed linear polarization of H$_2$O and methanol masers. And, finally, the detection of the
Zeeman counterpart F1 the spot F, if confirmed, is a definite proof of the interpretation of the 
linearly
polarized emission features as $\sigma$--components of Zeeman pattern.

\subsection{The nature of maser spots}

We have mapped 14 maser spots in W75N (Fig. 5), with two or possibly three coinciding Zeeman 
components.
Every maser spot has its own radial velocity, different from the radial velocity of other spots, and a
narrow line width of about 0.2 -- 0.3~km\,s$^{-1}$. 1667-MHz spots do not coincide in position with 
1665-MHz
spots. Almost every spot is 100 per cent elliptically polarized. We interpret them as $\sigma$--
components
of the Zeeman pattern. The magnetic field in the Zeeman pairs was determined to be from 5 to 11 
milligauss,
the same values as found in other OH masers \citep{garcia88}. One can assume that in 
other spots where the second
$\sigma$--component is not visible the magnetic field strength is of the same order.
If it is a $\sigma$--component, then the linear 
polarization
vector must be perpendicular to the direction of the magnetic field, and the
magnetic field  direction seems to be roughly the same for all spots.
This can be seen on the
histogram in Fig~6. Unfortunately the direction of the magnetic field 
can not be determined reliably from the linear polarization measurements
because of the uncertain amount of Faraday 
rotation in the Galaxy. The OH masers are located in the Galactic
disk where the Faraday rotation is at
maximum. Total Faraday rotation in the Galactic plane as measured
with extragalactic radio sources is

\begin{equation}
RM(l)=RM_0\sin(l_0-l) 
\end{equation}

where {\it RM$_0$}=1607~rad\,m$^{-2}$ $\pm$10~per cent r.m.s., {\it l$_0$}=62.1$^\circ$ \citep{clegg92}.
W75N is at the galactic longitude
{\it l}=81.9$^\circ$, and from (7) one has {\it RM}=--544~rad\,m$^{-2}$.
At the OH frequency ($\lambda=18$~cm) the Faraday rotation is (--544)$\times$(0.18)$^2$=--
17.6~rad$\pm$5.4~rad if one assumes maximum uncertainty 30 per cent.
This is the total Faraday rotation throughout the galactic disk; W75N is at the distance of 2 kpc, which is
probably about a half of the total effective distance, and the Faraday rotation to W75N can be a
factor of 2 lower, or --9~rad$\pm$2.7~rad. 
This is a large rotation, about 3 full turns, and a correction
for the Faraday rotation to the position angle of the linear polarization could be quite uncertain.
Therefore it is not posible to determine the direction of the magnetic field in OH maser spots.

	Physical parameters of the maser spots can be estimated from maser models, which require 
gas
density n$_{H_2}$=10$^7$cm$^{-3}$, kinetic temperature 100~K, dust temperature
150~K, and OH abundance 10$^{-5}$ \citep*{gray95}. Such parameters can
provide inversion of 1665~MHz OH transition in a model with FIR line overlap and a velocity
gradient of about 0.025~km\,s$^{-1}$/A.U. \citep*{gray95}. With the magnetic
field strength of 10 milligauss the model of \citet*{gray95} provides 100 per cent
elliptically polarized $\sigma$--components, with $\pi$--components suppressed,
in agreement with results of this paper for W75N. The size of maser spots 10 A.U. and
molecular hydrogen density 10$^7$cm$^{-3}$ correspond to the mass of maser spots
of 2$\times10^{-7}$M$_{\odot}$ , which is less than the mass of the Earth.

If the maser spots are discrete physical objects -- dense cold gas condensations
surrounded  a low density medium -- they should be confined
by the external pressure. A gas condensation with the density 10$^{7}$cm$^{-3}$
and temperature 100~K can be in pressure equilibrium with the gas of density
10$^{5}$cm$^{-3}$ and temperature 10$^4$~K.
However, the magnetic pressure in the maser spots, with the magnetic field
strength of 10 milligauss, is an order of magnitude higher, and can
be compensated by turbulent or ram pressure of the hot medium
(see discussion in a paper by \citet{reid87}).
Another model of maser spots proposed
for Class~II methanol masers \citep{slysh99} assumed that the maser spots 
are
extended gaseous envelopes of solid icy planets orbiting around O, B-stars, outside
their HII regions. OH molecules as well as methanol molecules are continuosly
supplied to the envelope by evaporation of ice from the surface of the planets.
In W75N the ultracompact HII region VLA1 (Fig.~5) marks the position of the central star
with luminosity 1.4$\times10^{5}$L$_{\odot}$ \citep{moore91} which corresponds to the
main sequence O9--star with mass 20M$_{\odot}$. The largest distance from VLA1 to
a maser spot is about 1000 mas, or 2000 astronomical units. At this distance from
a 20M$_{\odot}$ star the orbital velocity is 3~km\,s$^{-1}$ which is consistent with the observed
velocity range of maser spots in W75N. In this model the magnetic
field originates in the planets.

\subsection{The model of region W75N}

OH maser spots in W75N are located around the ultracompact HII region VLA1, 
with
spots J and K possibly associated with another ultracompact HII region, VLA2 
(Fig~5).
One model of OH masers places them at D-type ionization front
\citep*{elitzur78}
or in the photo-dissociation region \citep*{hartquist91} surrounding
the HII-region. In W75N the geometry of OH maser spots is consistent with these
models although the observed gradient of the radial velocity along the chain
of maser spots from about 10~km\,s$^{-1}$ at spots A, B to 3.7~km\,s$^{-1}$ at
spot I is not predicted by the models. This gradient is better represented in
the model of jet or in disk models. \citet{torrelles97} based on the elongated shape
of the ultracompact HII-region VLA1 and orientation of chains of H$_2$O and OH
maser spots parallel to the direction of the bipolar outflow, observed in W75N
on a much larger scale, suggested that the masers trace the outflow at scales
of about 1 arcsec and that VLA1 is the powering source of the molecular outflow.
Our more accurate measurements of the absolute position of OH masers show
that they are not projected at VLA1 (Fig~7 in \citet{torrelles97}) but are
displaced to form an arc arround VLA1. The H$_2$O masers seem to form an arc
which is closer to VLA1.  Such a geometry is better described by
a disk arround VLA1 in which OH maser spots are located at a distance
of about 2000 astronomical units from VLA1, and H$_2$O masers are located a factor
of 2 to 5 closer. The disk model was first suggested by \citet{haschick81}.
VLA1 marks the position of the ionization source and the center
of gravity of the system. As was suggested in Section 4.2 there
might be a 20M$_{\odot}$ O9-star at the
center of VLA1, which gravitationally holds the disk of radius 2000 astronomical units.
In this model the elongated shape of VLA1 reflects the distribution of disk material,
which became visible in radio due to the ionization by the central star. 
We can not determine direction of the magnetic field in the disk from linear polarization
data because of the  uncertain amount of Galactic Faraday rotation (see Section 4.2) but
we can state that the distribution of the magnetic field orientation is not random
and is well organized in some direction, as is evident from the
histogram of Fig~6.

\section{Conclusions}
\begin{itemize}
\item One of the results of this study is the discovery of several linearly polarized
maser spots with almost 100 per cent polarization.

\item For the first time polarization mapping of W75N in another main line OH transition,
1667 MHz was done.

\item We suggest that the OH masers are located in a thin
disk rotating around an O-star
which is the exciting star of the ultracompact HII region VLA1.
The diameter of the disk is 4000 A.U. Two separate maser
spots are connected with the ultracompact
HII region VLA2.

\item Spots A and B at 1665~MHz and M and O at 1667~MHz come from the same
positions. The polarization is oppositely circular, and the position difference
is very small. Based on these properties one can identify these two pairs as
$\sigma{\pm}$--components of Zeeman patten. Both pairs are elliptically polarized,
with nearly the same position angles of the polarization vector, exactly as
required for $\sigma$--components. Two maser spots show Zeeman pairs corresponding
to the magnetic field 5.2 mgauss and 7.7 mgauss.

\item The linearly polarized maser spots are shown to be $\sigma$--components in
the case of the magnetic field perpendicular to the line of sight. We 
tentatively detected a Zeeman pair consisting of the linearly polarized 
components, with a deduced magnetic field of 10.7 mgauss. The direction of 
the magnetic field as determined for linearly polarized spots is perpendicular
to the plane of the disk, not corrected for the Galactic Faraday rotation.
\end{itemize}

\begin{center}
\bf{Acknowledgements}
\end{center}

NRAO is a facility of the National Science Foundation operated under a cooperative
agreement by Associated Universities. VIS and IEV acknowledge partial support
from INTAS (grant N97-11451) 
and Russian Foundation for Basic Research (grant N01-02-16902).

\clearpage

%\begin{table*}
\begin{deluxetable}{lrrrrrrrrrrrr}
%\tablenum{1}
\tabletypesize{\scriptsize}
\rotate
\tablewidth{0pt}
\tablecaption{Maser spots in W75N}
\tablehead{
%&&&&&\nl
\colhead{Spot}&\colhead{v$_{lsr}$}&\colhead{$I$}&\colhead{$V$}&\colhead{$Q$}&\colhead{$U$}&\colhead{
$m_C$}&\colhead{$m_L$}&\colhead{$\chi$}&\colhead{$R.A.$}&\colhead{$Dec.$}&\colhead{Size}&\colhead{$\Theta$}\nl 
&\colhead{km\,s$^{-
1}$}&\colhead{Jy/beam}&\colhead{Jy/beam}&\colhead{Jy/beam}&\colhead{Jy/beam}&\colhead{per 
cent}&\colhead{per 
cent}&\colhead{$^\circ$}&\multicolumn{2}{c}{mas}&\colhead{mas}&\colhead{$^\circ$} }

\startdata
\multicolumn{13}{c}{\bf{}1665 MHz}\nl

A & 12.45 & 16.88 &   16.24 & $<$0.04 & $-$1.56 &     94.4 &   9.3 & $-$45.0 & $-$301.7 & $-
$1176.7 &  5.9$\times$1.4 &  95\nl
B &  9.38 &  3.92 & $-$3.24 & $-$0.17 & $-$0.68 &  $-$84.4 &  17.9 & $-$52.2 & $-$301.1 & $-
$1177.3 &  5.5$\times$4.6 &  86\nl
C &  9.35 &  3.27 &    3.50 &    0.11 & $-$0.19 &    105.3 &   6.9 & $-$29.8 &  $-$98.2 &  $-$573.5 
&  5.6$\times$4.1 &  61\nl
D &  7.35 &  0.61 &    0.62 & $<$0.02 & $<$0.02 &    101.2 &$<$5.0 &    \dots&  $-$1.0 &   $-$32.7 
&  4.4$\times$1.1 &  83\nl
E &  6.00 &  5.76 & $<$0.02 &    4.71 &    2.69 &   $<$0.3 &  94.2 &    14.9 & $-$122.7 &  $-$663.6 
&  5.9$\times$2.2 &  92\nl
F $^{*}$ &  5.65 & 10.16 & $-$0.81 &    6.93 &    7.12 &   $-$9.8 &  97.8 &    22.9 &      0.0 & 0.0 &  
5.4$\times$1.2 &  96\nl
G &  5.29 &  3.30 & $-$0.78 &    2.37 &    1.87 &  $-$25.4 &  91.4 &    19.2 &    155.5 &     114.2 &  
7.6$\times$2.9 &  59\nl
H &  5.11 &  2.86 & $-$2.24 &    1.41 & $-$1.45 &  $-$80.0 &  70.7 & $-$22.9 &    186.9 &     167.1 
&  5.1$\times$2.5 &  94\nl
I &  3.70 &  1.18 & $-$1.28 & $<$0.02 &    0.13 & $-$110.8 &  10.8 &    40.6 &    249.0 &     317.9 
&  6.5$\times$2.9 &  77\nl
J &  3.00 &  8.16 &    0.79 &    7.61 & $-$3.05 &      7.9 & 100.5 & $-$10.9 &    299.8 & $-$1275.8 
& 10.7$\times$2.4 & 114\nl
K &  0.65 &  0.83 &    0.22 &    0.78 & $<$0.009&     24.2 &  93.7 &     0.0 &    376.5 & $-$1385.3 
& 19.7$\times$11.9& 142\nl
F1& -0.65 &  0.040 & $<0.003$ &  0.035 & 0.023&    $<10$ &  87   &    20.5 &     0.14 &      0.47 &  
6.5$\times$0.7&   74\nl
\multicolumn{13}{c}{\bf{}1667 MHz}\nl
L & 9.80 &  21.33 & $-$4.46 &    6.01 &    6.54 &  $-$22.7 &  41.6 &    23.7 & $-$175.4 &  $-$661.2 
&  6.2$\times$1.9 &  91\nl
M & 8.65 &   1.49 &    1.26 & $<$0.01 &    0.14 &     82.5 &   9.4 &    45.0 & $-$149.9 &  $-$506.1 
&  5.4$\times$0.7 &  90\nl
N & 6.55 &   0.55 & $-$0.61 &    0.045& $<$0.01 & $-$112.8 &   8.5 &     8.0 & $-$136.2 &  $-
$526.8 & 14.2$\times$3.2 & 176\nl
O & 5.95 &   2.46 & $-$2.23 & $<$0.01 &    0.17 &  $-$92.5 &   7.1 &    45.0 & $-$149.7 &  $-
$505.0 &  7.4$\times$2.5 & 113\nl

\enddata
%\tableline
%\end{tabular}                       
\begin{list}{}{}
\item[$^*$]Reference position: R.A.=20$^{\rm h}$38$^{\rm m}$36$\fs$414$\pm$0$\fs$003 
Dec.=42$\degr$37$\arcmin$35$\farcs$44$\pm$0$\farcs$03 (J2000)
\end{list}
\end{deluxetable}
                        	       
\clearpage

\begin{figure}
\epsscale{0.30}
\plotone{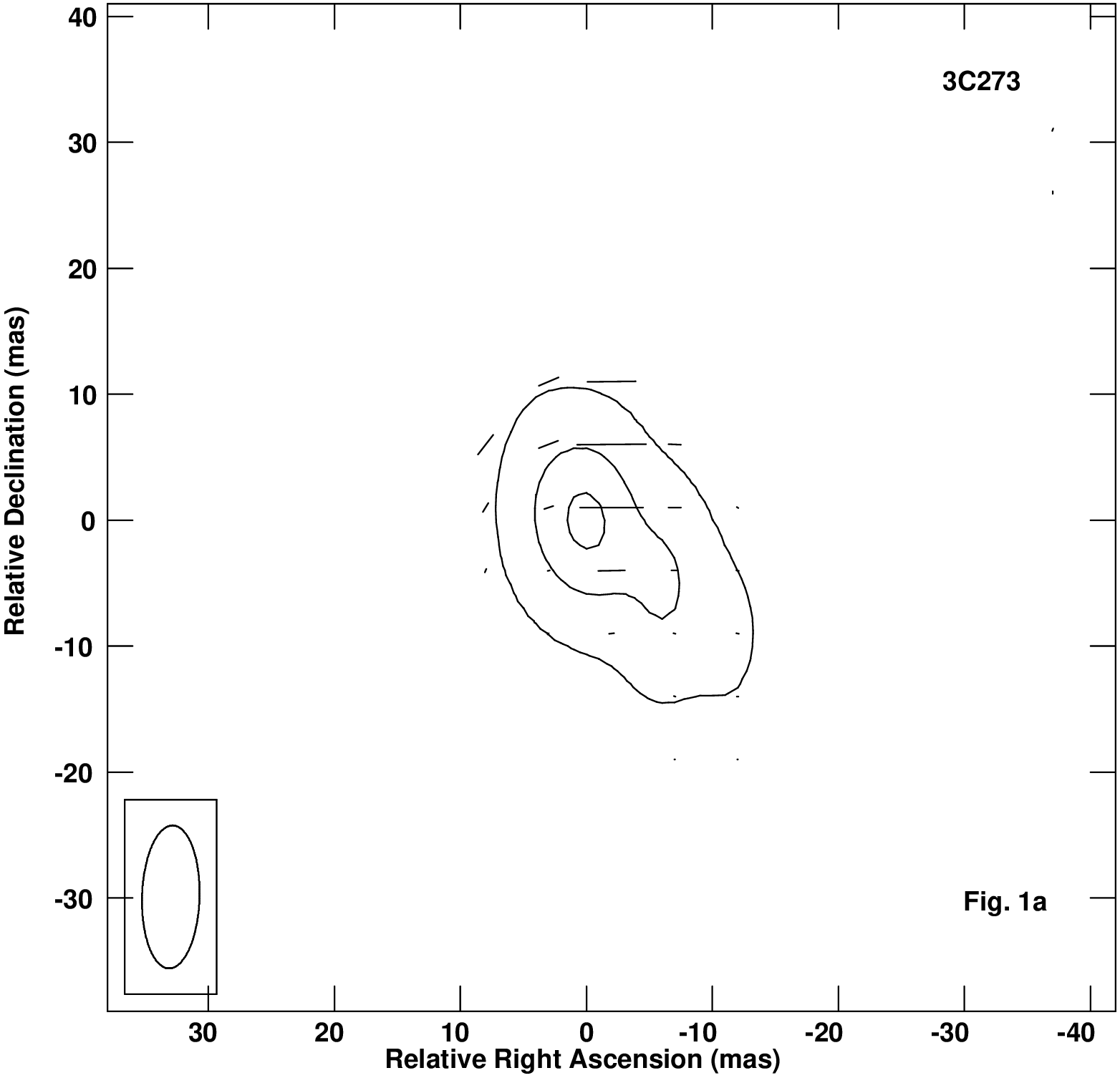}
\plotone{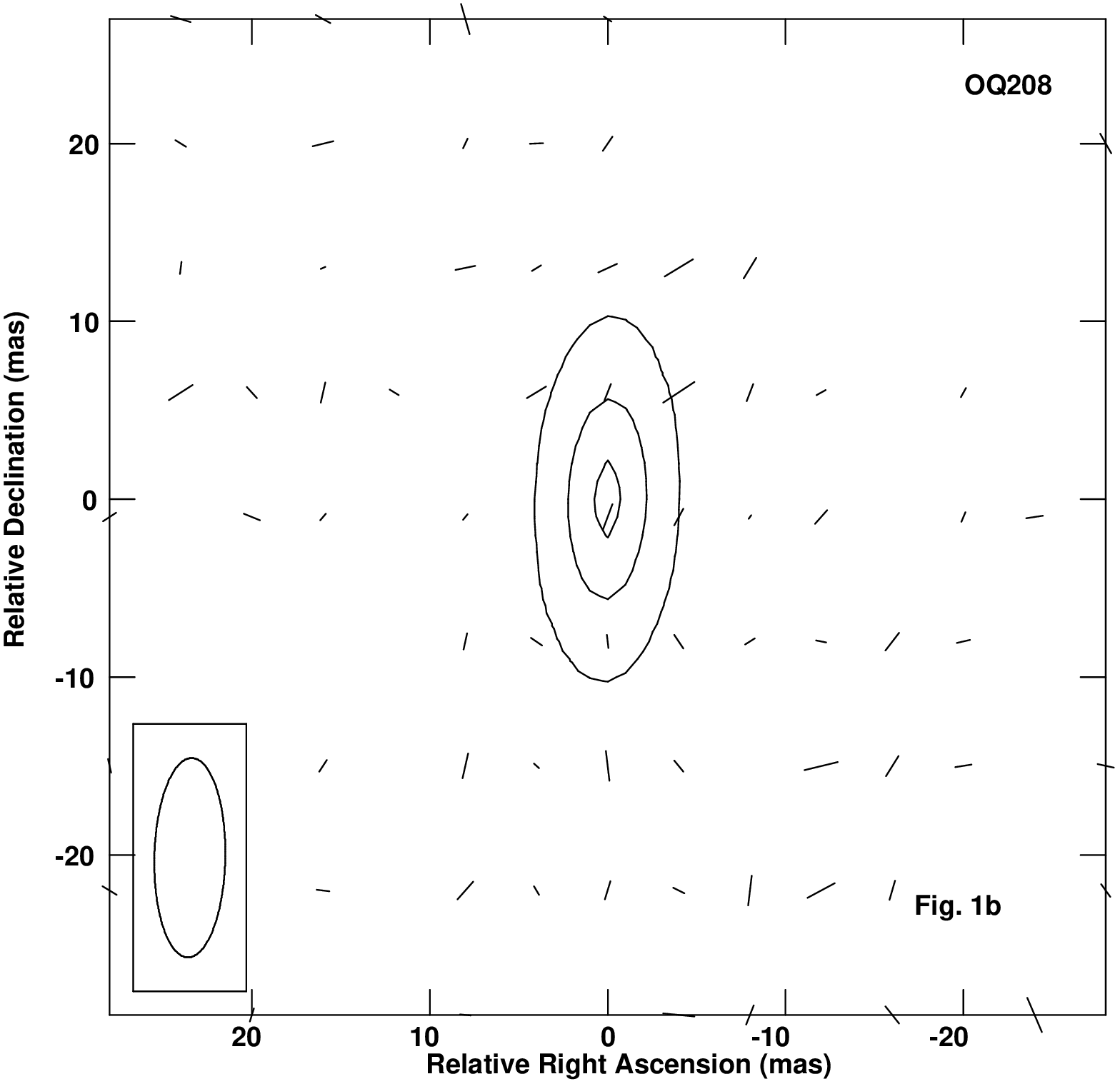}
\caption{Continuum maps of calibration sources.
a) 3C273.
Contours - total intensity I: 0.9, 4.6, 8.3 Jy/beam;
vectors - linearly polarized intensity, maximum length=0.125 Jy/beam;
m$_L=1.4\%$; m$_C=1.8\%$ 
b) OQ208.
Contours - total intensity I: 0.085, 0.45, 0.77 Jy/beam;
vectors - linearly polarized intensity; 1 mas=0.0004 Jy/beam.
m$_L\le0.8\%$; m$_C=1.8\%$.  }

\end{figure}

\clearpage

\begin{figure}
\epsscale{0.80}
\plotone{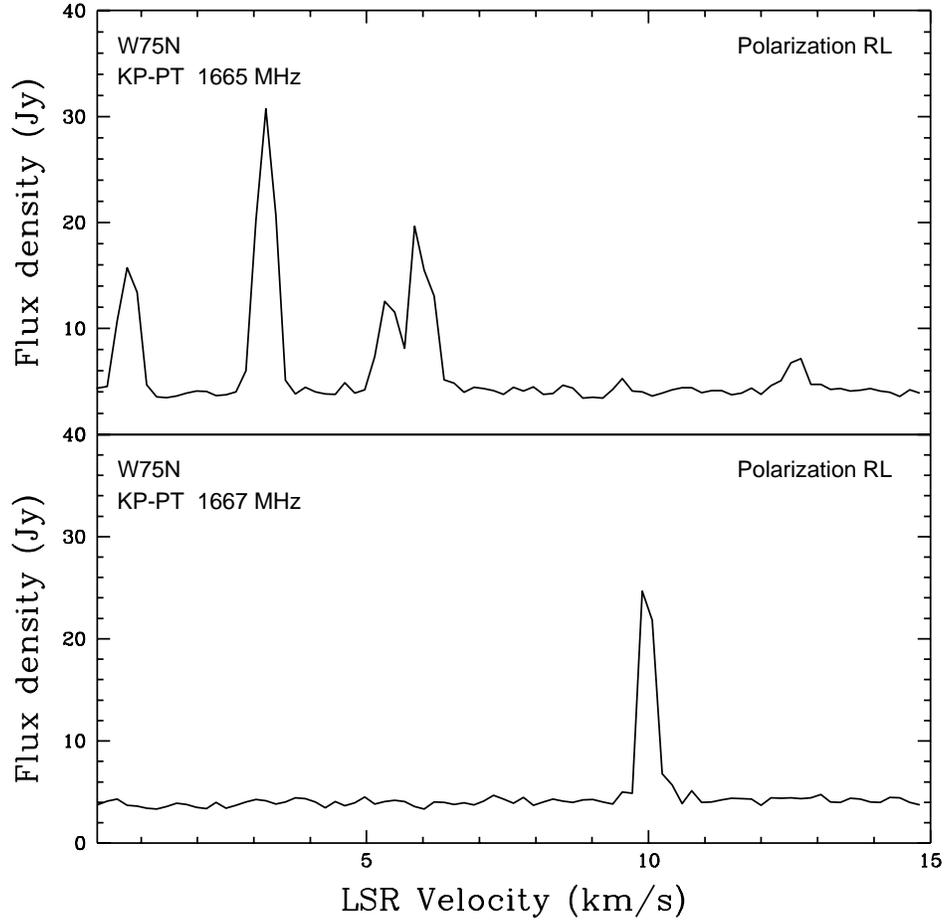}
\caption{ The cross power spectra of W75N on the base line Kitt Peak - Pie Town.
Upper: cross-polarized RL spectrum of 1665~MHz OH line;
lower: cross-polarized RL spectrum of 1667~MHz OH line. }

\end{figure}

\clearpage

\begin{figure}
\epsscale{0.80}
\plotone{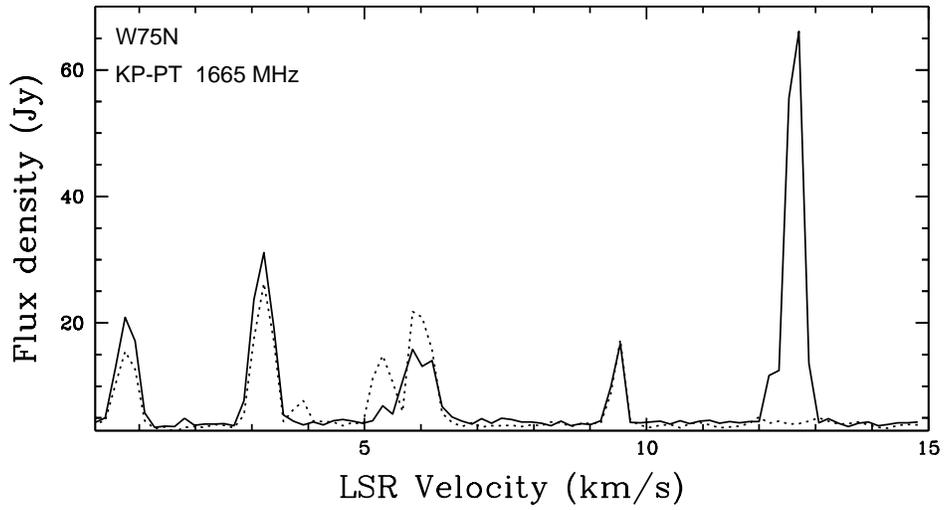}
\caption{The cross power circularly polarized spectra of W75N on the base line
Kitt Peak - Pie Town in 1665~MHz OH line.
Dotted line: left-circular polarization (LCP);
solid line: right-circular polarization (RCP).}

\end{figure}

\clearpage

\begin{figure}
\epsscale{0.80}
\plotone{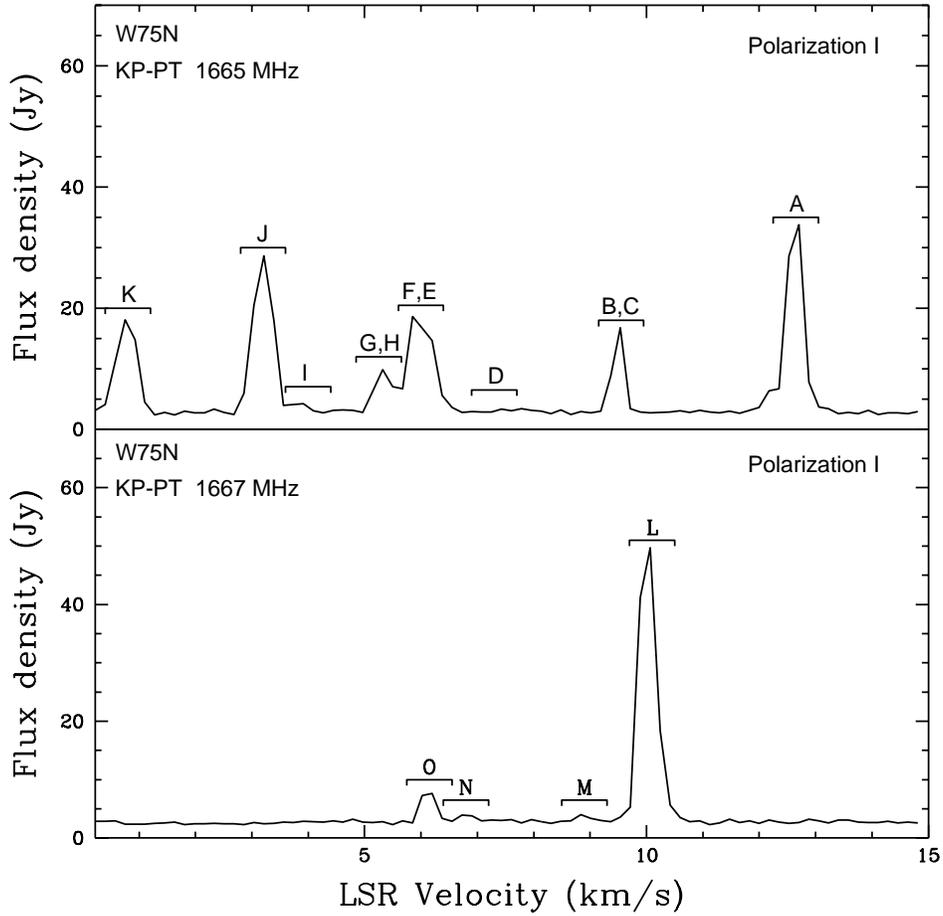}
\caption{Total intensity I cross power spectra of W75N in 1665~MHz (upper) and
1667~MHz (lower) OH lines.
Letters A - O mark spectral features which were mapped and their map and polarization
parameters are given in Table 1.} 
\end{figure}

\clearpage

\begin{figure}
\epsscale{0.80}
\plotone{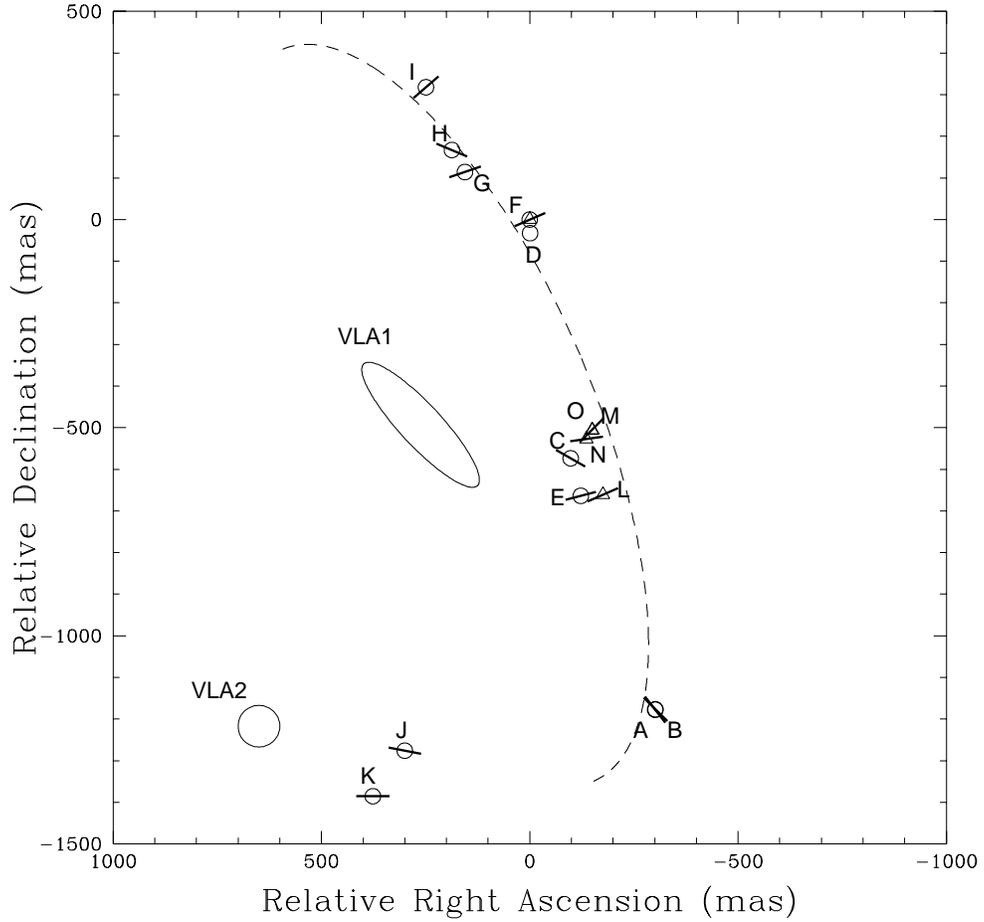}
\caption{ A VLBA map of the OH maser features. Circles denote
1665~MHz features, triangles - 1667~MHz features. They are labeled as in Table 1.
Vectors attached to the maser features show relative direction of the
magnetic field. An ellipse and a circle labelled VLA1 and VLA2 show position and size of ultra compact
HII regions, from Torrelles et al (1997). Dashed line shows a portion of a possible disk centered
at VLA1 and inclined by 12$^\circ$ to the line of sight.}

\end{figure}

\clearpage

\begin{figure}
\epsscale{0.80}
\plotone{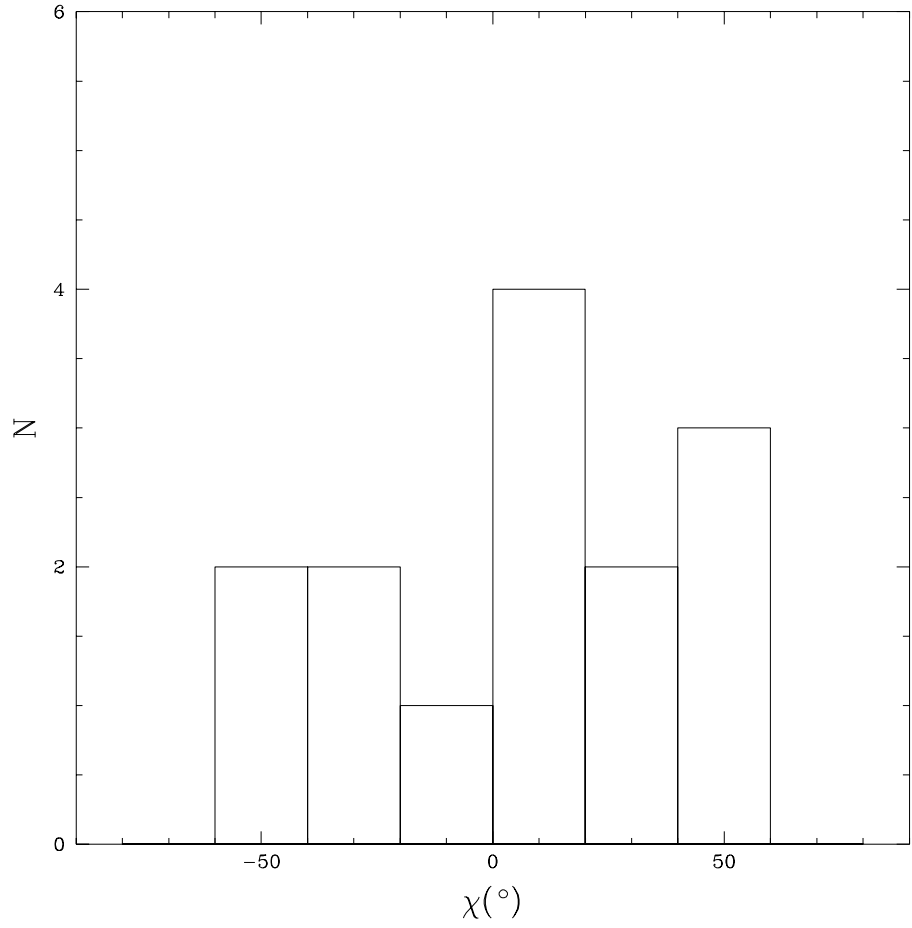}
\caption{Distribution of the linear polarization relative position angles $\chi$ for maser spots
in W75N.  $\bar{\chi}=4^\circ\pm17^\circ$.  }

\end{figure}  

\clearpage

\begin{figure}
\epsscale{0.30}
\plotone{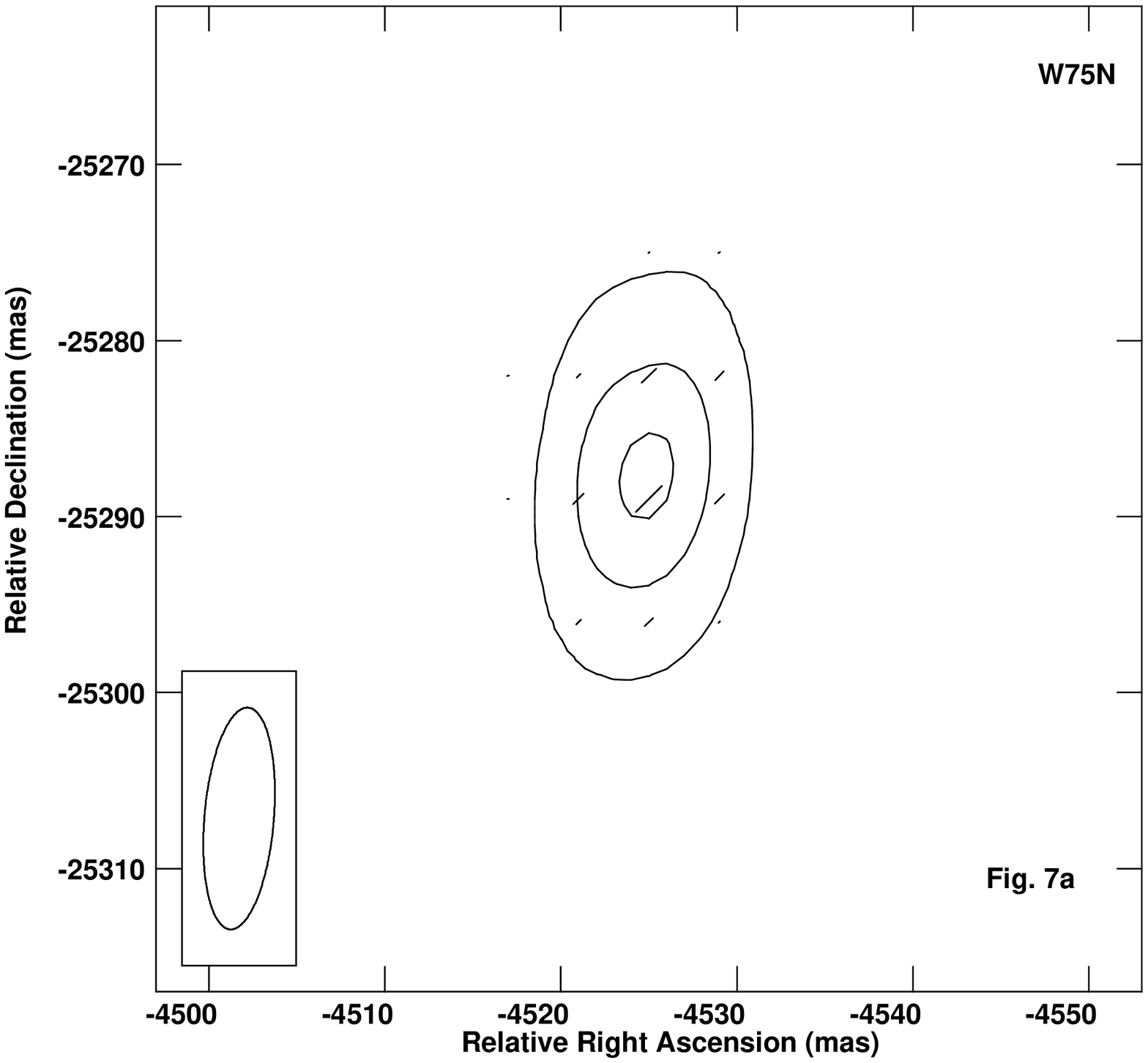}
\plotone{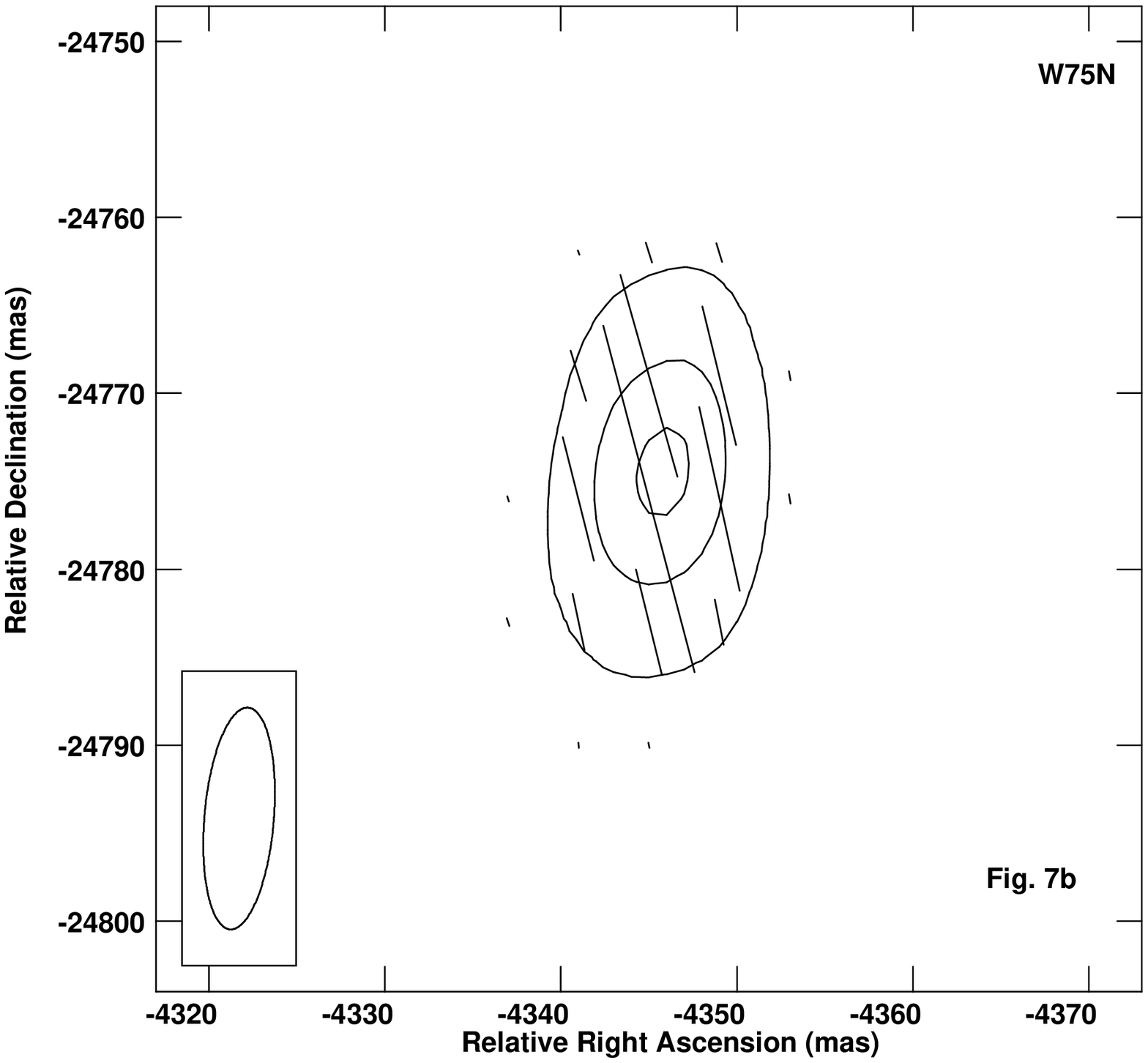}
\plotone{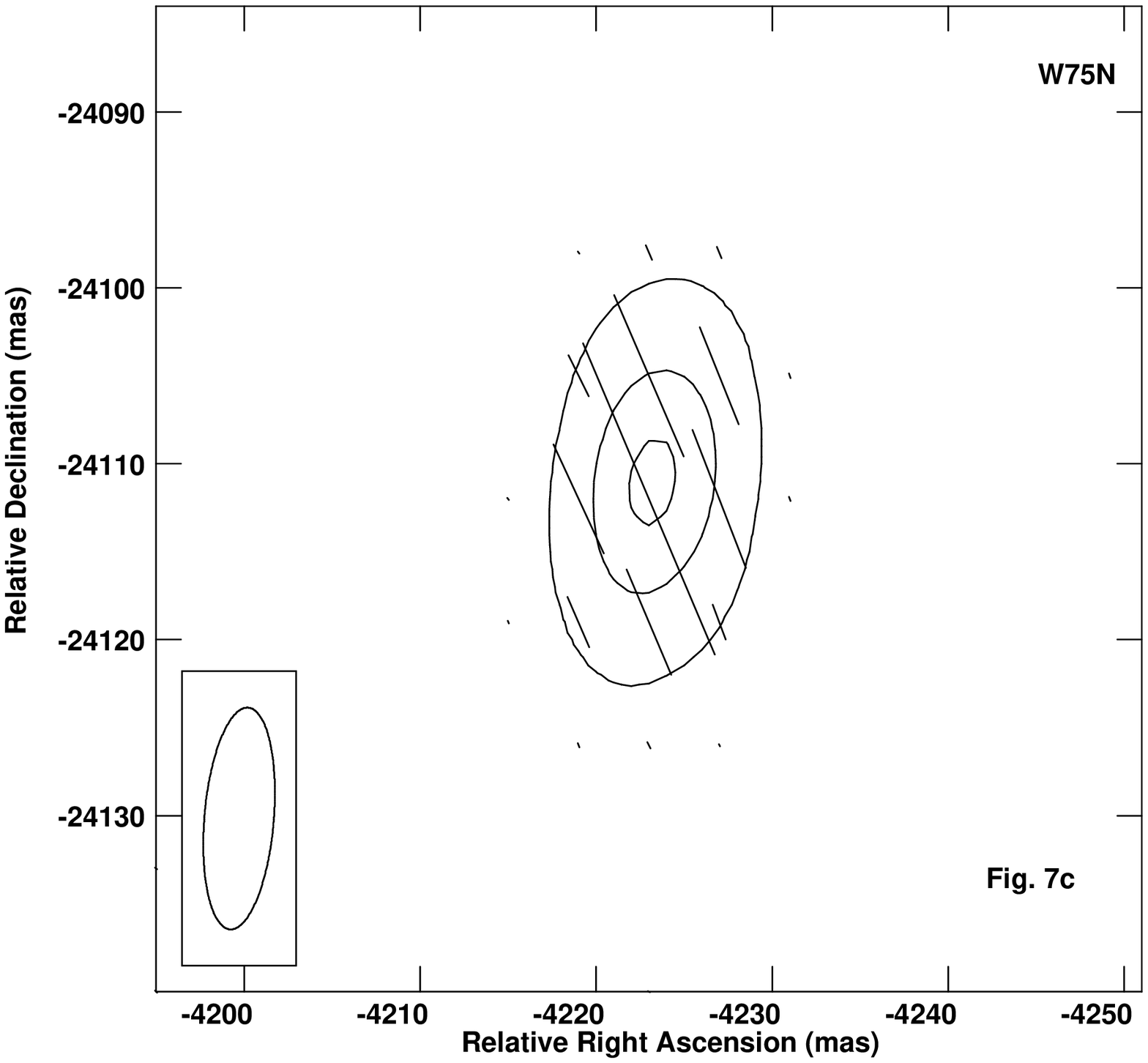}
\plotone{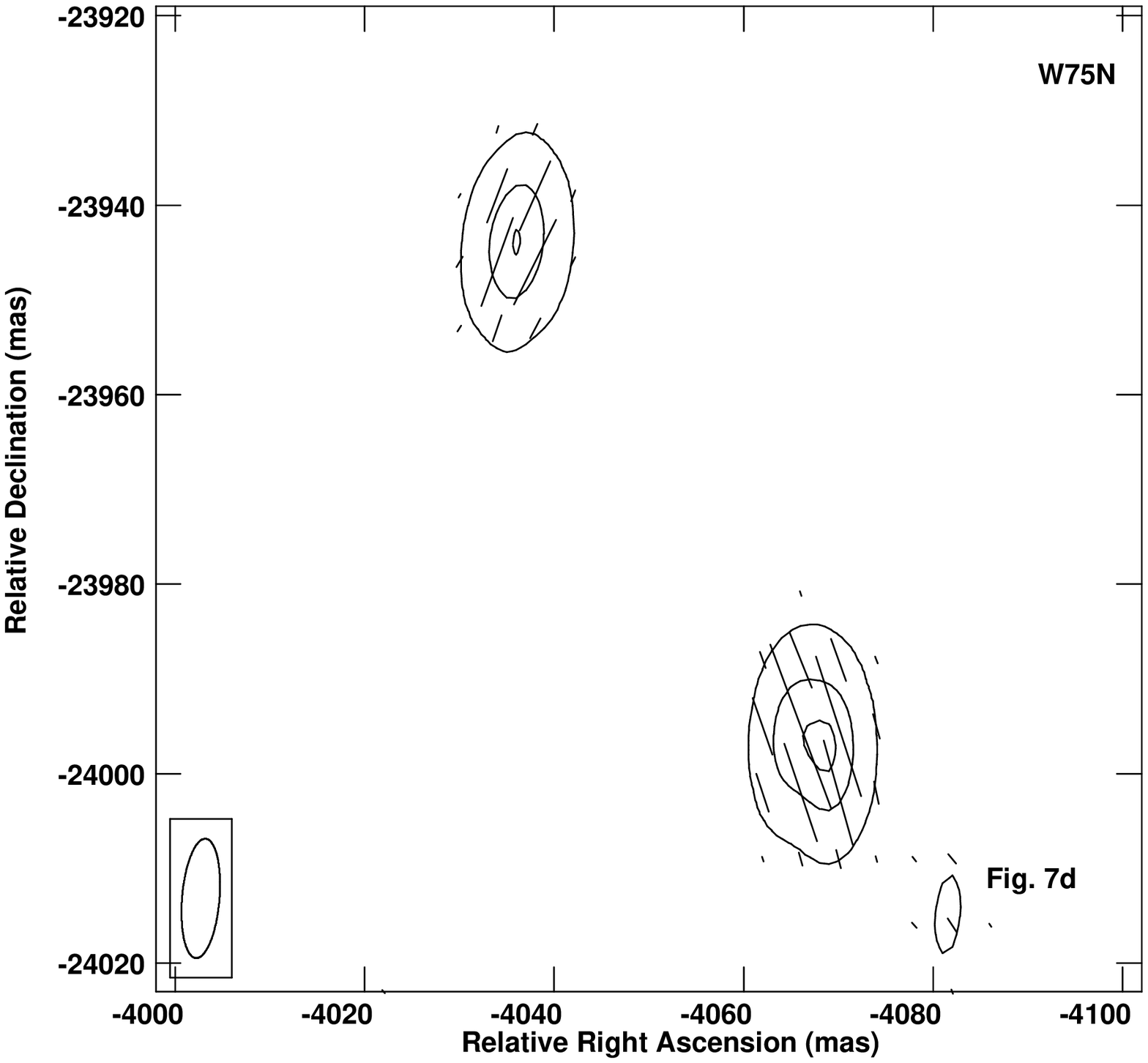}
\plotone{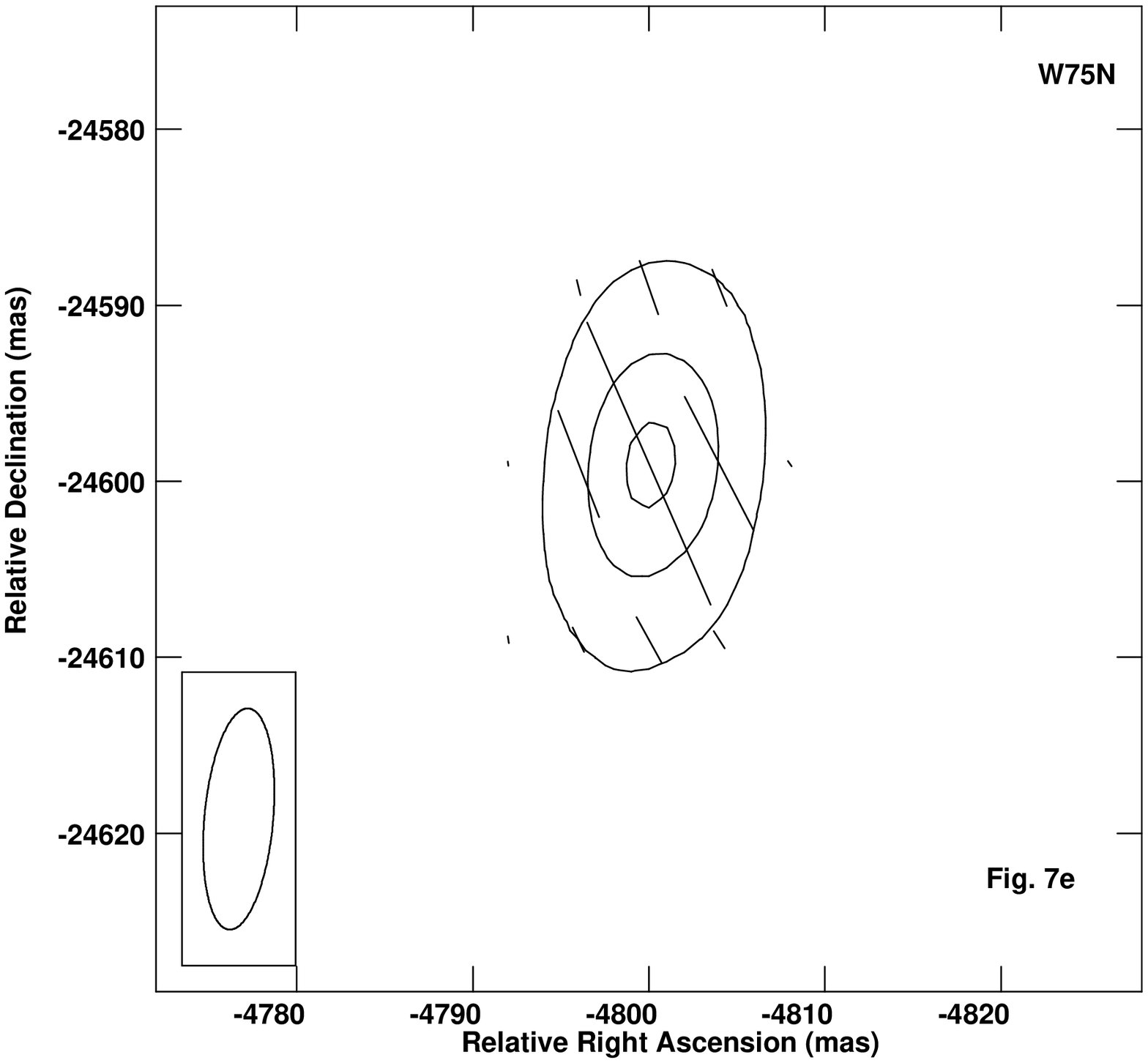}
\caption{Linearly polarized and total intensity - maps of maser spots in W75N.
a) Spot A: vector - linear polarization intensity (length) and relative position angle (direction);
contours -- total intensity: 0.1, 0.5, 0.9 $\times$ (peak intensity).
Peak values $\sqrt{Q^2+U^2}=1.6$ Jy/beam, I=16.9 Jy/beam; 								 
b) Spot E: $\sqrt{Q^2+U^2}=5.4$ Jy/beam, I=5.8 Jy/beam;
c) Spot F: $\sqrt{Q^2+U^2}=9.9$ Jy/beam, I=10.2 Jy/beam;
d) Spots G and H: (G) $\sqrt{Q^2+U^2}=3.0$ Jy/beam, I=3.3 Jy/beam;
		(H) $\sqrt{Q^2+U^2}=2.0$ Jy/beam, I=2.9 Jy/beam;
e) Spot L: $\sqrt{Q^2+U^2}=8.9$ Jy/beam, I=21.3 Jy/beam. }
\end{figure}

\clearpage

\begin{figure}
\epsscale{0.80}
\plotone{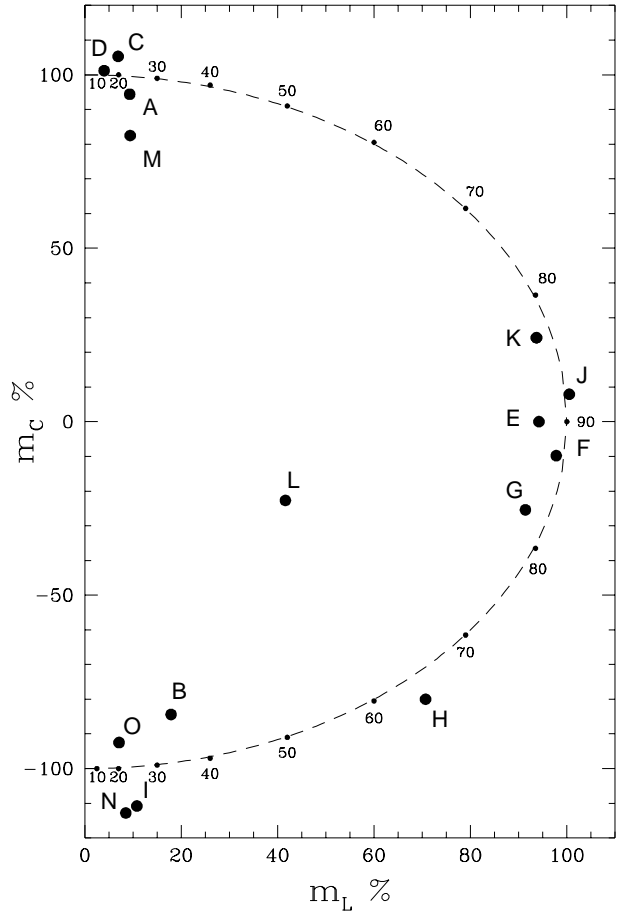}
\caption{ Percentage of circular polarization m$_C$ of maser spots in W75N, plotted
against percentage of linear polarization m$_L$. A completely polarized feature
would lie on the dotted line. Negative values of m$_C$ indicate LH circular or elliptical
polarization and correspond to negative V in Table~1.   } 
\end{figure} 

\clearpage

\begin{figure}
\epsscale{0.80}
\plotone{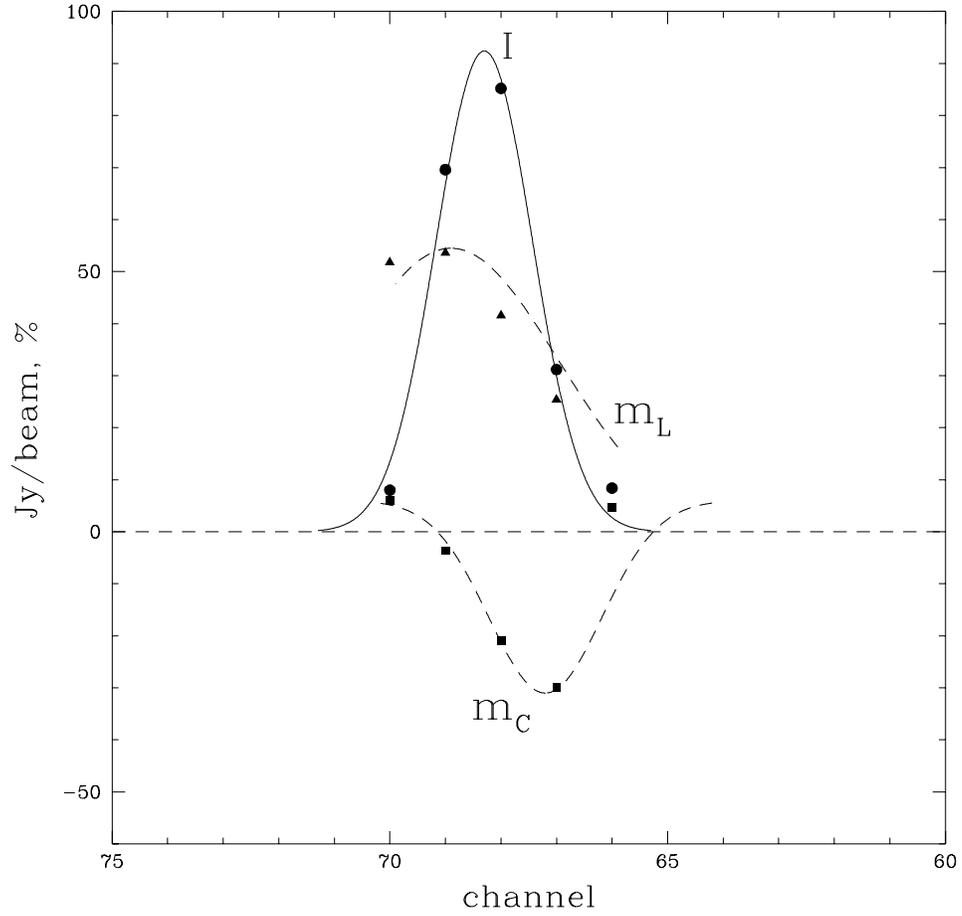}
\caption{ Line profile of 1667~MHz Spot L. Solid line (filled circles): total
intensity I; dashed line (triangles): percentage of linear polarization
m$_L$; dashed line (squares): percentage of circular polarization m$_C$.  } 
\end{figure}


\begin{thebibliography}{}
\bibitem[Baart et al.(1986)]{baart86} Baart, E. E., Cohen, R. J., Davies, R. D., Norris, 
R. P., and Rowland, P. R. 1986, \mnras, 219, 145
\bibitem[Clegg et al.(1986)]{clegg92} Clegg, A. W., Cordes, J. M., Simonetti, J. K., and
Kulkarni, S. R.  1992, \apj, 386, 143
\bibitem[Conway et al.(1993)]{conway93} Conway, R. G., Garrington, S. T., Perley, 
R. A., and Biretta, J. A. 1993, \aap, 267, 347
\bibitem[Cook(1966)]{cook66} Cook, A. H., 1966 \nat, 211, 503 
\bibitem[Davies et al.(1977)]{davies77} Davies, R. D., Booth, R. S., and Perbet, J.-N.
1977, \mnras, 181, 83 
\bibitem[Elitzur(1996)]{elitzur96} Elitzur, M. 1996, \apj, 457, 415
\bibitem[Elitzur and de Jong(1978)]{elitzur78} Elitzur, M., and de Jong, T. 1978, \aap, 67, 323 
\bibitem[Garcia-Barreto et al.(1988)]{garcia88} Garcia-Barreto, J. A., 
Burke, B. F., Reid, M. J., Moran, J. M., Haschick, A. D., and
Schilizzi, R. T. 1988, \apj, 326, 954  
\bibitem[Goldreich, Keeley, and Kwan(1973)]{goldreich73} Goldreich, P., Keeley, D. A., and Kwan, 
J. Y. 1973, \apj, 179, 111
\bibitem[Gray and Field(1995)]{gray95} Gray, M. D., and Field, D. 1995, \aap, 298, 243     
\bibitem[Habing et al.(1974)]{habing74} Habing, H. J., Goss, W. M., Mattews, H. E., and
Winnberg, A. 1974, \aap, 35, 1
\bibitem[Hartquist and Sternberg(1991)]{hartquist91} Hartquist, T. W., and Sternberg, A., 1991, 
\mnras, 248, 48      
\bibitem[Haschick et al.(1981)]{haschick81} Haschick, A. D., Reid, M. J., Burke, 
B. F., Moran, J. M., and Miller, G. 1981, \apj, 244, 76 
\bibitem[Hutawarakorn and Cohen(1999)]{hutawarakorn99}Hutawarakorn, B. and Cohen, R. J., 1999, \mnras, 303, 845 
\bibitem[Migenes et al.(2001)]{migenes01} Migenes, V., 
Slysh, V. I., Val'tts, I. E., Horiuchi, S., and Inoue, M. 2001, \apj, (in 
preparation)
\bibitem[Moore et al.(1991)]{moore91} Moore, T. J. T., 
Mountain, G. M., and Yamashita, T. 1991, \mnras, 248, 79
\bibitem[Reid et al.(1987)]{reid87} Reid, M. J., 
Myers, P. C., and Bieging, J. H. 1987, \apj, 312, 830
\bibitem[Slysh et al.(1999)]{slysh99} Slysh, V. I., Val'tts, I. E., Kalensky, S. V., and
Larionov, G. M. 
1999, Astronomy Report, 43, 785
\bibitem[Stanghelini et al.(1998)]{stanghelini98} Stanghelini, C., Dallacasa, D., 
O'Dea, C. P., Baum, S. A., Fanti, C., and Fanti, R. 
1998, Proceedings of the IAU Colloquium 164, eds. J. A. Zensus, G. B. Taylor,
\& J. M. Wrobel, A.S.P. Conference Series, p.177  
\bibitem[Torrelles et al.(1997)]{torrelles97} Torrelles, J. M., Gomez, J. F., Rodriguez, L. F., 
Ho, P. T. P., Curiel, S., and Vazquez, R.
1997, \apj, 489, 744
\end{thebibliography}
\end{document}